\documentclass[letterpaper,english,reprint, aps]{revtex4-1}
\usepackage[T1]{fontenc}
\usepackage[latin9]{inputenc}
\usepackage{color}
\usepackage{mathrsfs}
\usepackage{amsmath}
\usepackage{amssymb}
\usepackage{graphicx}
\usepackage{esint}

\makeatletter

\pdfpageheight\paperheight
\pdfpagewidth\paperwidth

\makeatother

\usepackage{babel}
\begin{document}
\title{Performance optimization of Nernst-based thermionic engines}
\author{Wei Yan$^{1}$}
\thanks{The two authors contributed equally to this work.}
\author{Minglong Lv$^{1}$}
\thanks{The two authors contributed equally to this work.}
\author{Ousi Pan$^{1}$}
\author{Zhimin Yang$^{2}$}
\author{Jincan Chen$^{1}$}
\author{Shanhe Su$^{1}$}
\email{sushanhe@xmu.edu.cn}

\address{$^{1}$Department of Physics, Xiamen University, Xiamen 361005, People's
Republic of China~\\
$^{2}$School of Physics and Electronic Information, Yan\textquoteright an
University, Yan\textquoteright an 716000, People\textquoteright s
Republic of China}
\date{\today}
\begin{abstract}
In this paper, we examine the power and efficiency of the thermionic
device utilizing the Nernst effect, with a specific focus on its potential
application as an engine. The device operates by utilizing the vertical
heat current to generate a horizontal particle current against the
chemical potential. By considering the influence of a strong magnetic
field, we derive analytical expressions for the current and heat flux.
These expressions are dependent on the temperature and chemical potential
of heat reservoirs, providing valuable insights into the device performance.
The impact of driving temperatures on the performance of the thermionic
engine has been assessed through numerical analysis. The research
findings will guide the experimental design of Nernst-based thermionic
engines.
\end{abstract}
\maketitle
The Nernst effect refers to a thermoelectric or thermomagnetic phenomenon
that is observed in electrically conductive materials when subjected
to perpendicular magnetic field and temperature gradient \citep{goldsmid2010introduction,behnia2016nernst,chiang2024nernst}.
This phenomenon is a result of charge carriers diffusing in response
to the magnetic field, generating a transverse electric field that
is directly proportional to the applied temperature gradient. 

The Nernst effect has primarily been studied and observed in metallic
and semiconducting materials \citep{price1956theory,silverman1963nernst,watzman2018dirac,murata2020enhancement,masuki2021origin}.
It has applications in various areas, such as thermoelectric devices,
spintronics, and energy harvesting. Sothmann theoretically proposed
Nernst engines based on quantum Hall edge states, where they are identified
to have the performance surpassing classical counterparts \citep{sothmann2014quantum}.
Graphene, with its distinctive electronic and thermal properties,
has also garnered significant attention in the investigation of the
Nernst effect \citep{checkelsky2009thermopower}. The presence of
a magnetic field perpendicular to the graphene sheet can give rise
to intriguing transport phenomena attributable to the quantum Hall
effect and the Landau quantization of electronic states. Bergman proposed
a theory of conductivity that is expressed in terms of entropy per
carrier, offering valuable insights into the characteristics of Nernst
thermopower in two-dimensional graphene materials \citep{bergman2010theory}.
Sharapov provided an insightful visualization of the Nernst effect
in Laughlin geometry by employing an ideal reversible thermodynamic
cycle \citep{sharapov2021nernst}. Investigation on the reduction
of magnetic field intensity has revealed an enhanced spin Nernst effect,
which demonstrates sensitivity to both sample characteristics and
contacts \citep{cheng2008spin}. In addition, the unique characteristic
of the anomalous Nernst effect, which does not depend on a strong
magnetic field, has attracted significant attention in various ferromagnetic
materials \citep{yang2020giant,sakai2018giant,caglieris2018anomalous,ikhlas2017large,tang2024unconventional,kurosawa2024large}.

It is crucial to emphasize that the practical realization of the classical
Nernst engine faces significant challenges that need to be addressed
\citep{brandner2013multi,brandner2013strong}. Firstly, the existing
devices suffer from low efficiency, which severely limits their power
generation capability. Moreover, the implementation of magnetic fields
in practical settings requires some cost. Under these circumstances,
a structure for generating thermoelectric energy via the ordinary
Nernst effect in the absence of an external magnetic field has been
proposed \citep{murata2024zero}. A simplified model has been proposed
for an engine that harnesses the Nernst effect. This model revolves
around the migration of electrons between four heat reservoirs operating
at different temperatures \citep*{stark2014classical,Brandner2015UniversalBO}and
encompasses the transport of heat and particles in non-interacting
systems, drawing an analogy to the Landauer-Büttiker approach \citep{pastawski1991classical}.
In this study, we extensively delve into the theoretical framework,
providing comprehensive expressions for current and heat flow within
the classical Nernst engine. Furthermore, we conduct thorough calculations
to determine the power and efficiency of the Nernst engine under various
conditions, thereby revealing the optimal performance and associated
parameters.

Figure 1 depicts the geometric configuration of the Nernst-based thermionic
engine, which comprises a circular two-dimensional central region
positioned perpendicular to a uniform magnetic field with a magnitude
of \textcolor{black}{$B$}. \textcolor{black}{The central region possesses
a radius of $R$} and is surrounded by four distinct thermochemical
reservoirs. Electrons in reservoir $C_{i}$ is characterized by the
chemical potential $\mu_{i}$ and temperature $T_{i}$. Each of these
reservoirs encompasses a segment of length $l$, which is equal to
$\pi R/2$.
\begin{figure}[h]
\includegraphics[scale=0.5]{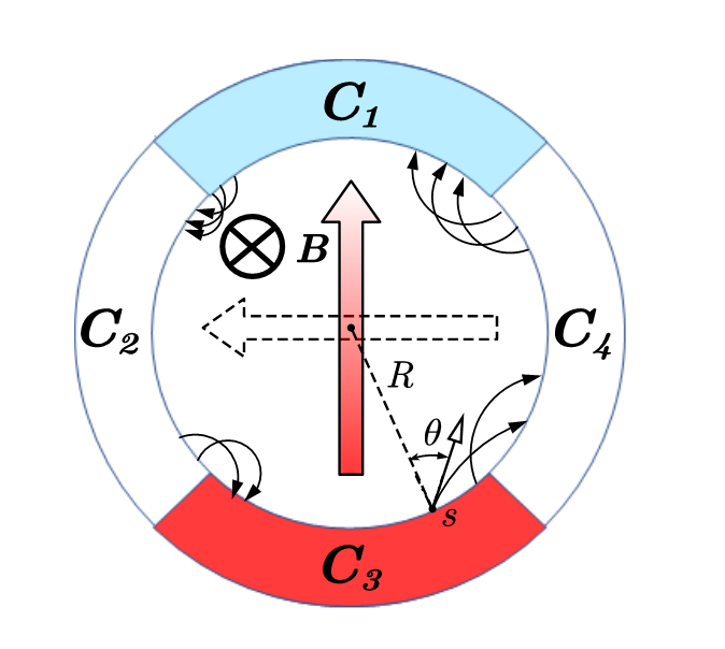}

\caption{The scheme diagram of a Nernst-based thermionic engine. Reservoir
$C_{3}$ possesses a higher temperature compared to reservoir $C_{1}$
( $T_{3}>T_{1}$), while reservoir $C_{2}$ has a higher chemical
potential than reservoir $C_{4}$ ($\mu_{2}>\mu_{4}$). The red gradient
arrow represents the flow of heat current, while dashed arrow represents
the movement direction of particles. The circular arrow denote a typical
trajectory for a electron under a strong magnetic field $B$. For
example, an electron may leave reservoir $C_{3}$ at \textcolor{black}{the
position $s$}\textcolor{red}{{} }with an angle $\theta$ and transports
to reservoir $C_{4}$. }
\end{figure}

When an electron reaches the circular boundary from one of the reservoirs,
it is assumed to enter the central region, where it undergoes a circular
trajectory due to the influence of the Lorentz force. The average
number of electrons with the range $\left(p_{r},p_{r}+dp_{r}\right)$
of radial momentum and $\left(p_{s},p_{s}+dp_{s}\right)$ of tangential
momentum, located in a small area $drds$ at the boundary of reservoir
$C_{i}$, is expressed as follows
\begin{equation}
dN_{i}\equiv2\exp\left[-\beta_{i}\left(E-\mu_{i}\right)\right]drdsdp_{r}dp_{s}/h^{2},
\end{equation}
where the approximation of Maxwell-Boltzmann statistics has been applied.
Here, $2$ denotes the spin of the electron, $r$ represents the radial
coordinate, and $E=\left(p_{r}^{2}+p_{s}^{2}\right)/(2m)$ is the
kinetic energy of the electrons with $m$ being the mass of electron,
$h$ is Planck\textquoteright s constant, and $\beta_{i}=1/\left(k_{B}T_{i}\right)$
with $k_{B}$ being Boltzmann's constant. For $p_{r}<0$, any particle
that contributs to $dN_{i}$ will reach the boundary within the time
interval $dt=-mdr/p_{r}$ . Through the elimination of $dr$ in favor
of $dt$ and the application of a change of variables $p_{r}=-\sqrt{2mE}\cos\theta$
and $p_{s}=\sqrt{2mE}\sin\theta$, we can express Eq. (1) in a different
form. This change of variables leads to the relation
\begin{equation}
dN_{i}/dt=\frac{2\sqrt{2mE}}{h^{2}}\exp\left[-\beta_{i}\left(E-\mu_{i}\right)\right]\cos\left(\theta\right)dsdEd\theta.
\end{equation}
By integrating over variables $s$, $E$, and $\vartheta$, the total
electron current $J_{i}^{+}$ flowing from the reservoir $C_{i}$
into the central region is given by

\begin{align}
J_{i}^{+} & =2\intop_{l}ds\intop_{0}^{\infty}dE\intop_{-\pi/2}^{\pi/2}d\theta\cos\left(\theta\right)u_{i}\left(E\right)\nonumber \\
 & =\frac{2\sqrt{2\pi m}l}{h^{2}\beta_{i}^{3/2}}e^{\beta_{i}\mu_{i}},
\end{align}
where $u_{i}\left(E\right)=\sqrt{2mE}\exp\left[-\beta_{i}\left(E-\mu_{i}\right)\right]/h^{2}$.

\begin{figure}[h]
\includegraphics[scale=0.4]{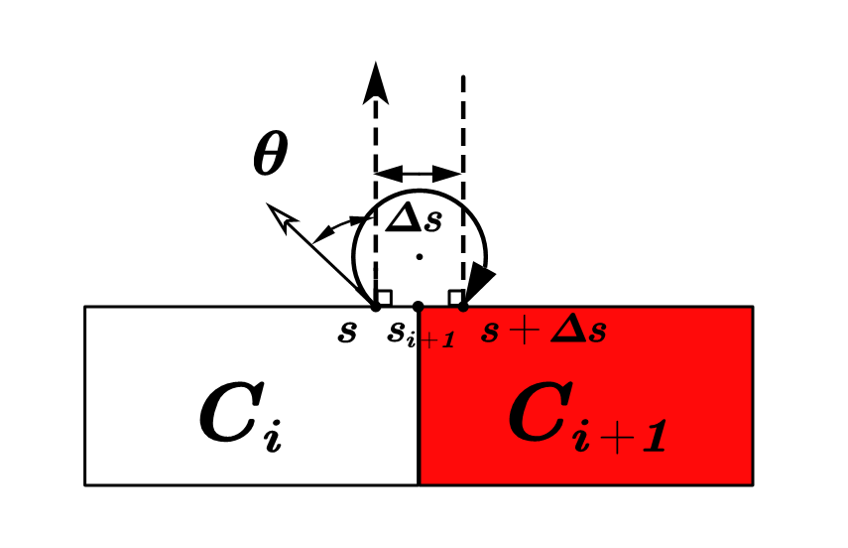}

\caption{The trajectory of an electron starts at reservoir $C_{i}$ from position
$s$ with an angle $\theta$, and enters reservoir $C_{i+1}$ at position
$s+\mathit{\Delta}s$.}
\end{figure}

By assuming that each electron reaching the boundary from the central
region is absorbed in the adjacent reservoir, the expression for the
steady-state current $J_{i}^{-}$ flowing into $C_{i}$ is calculated
as follows

\begin{equation}
J_{i}^{-}=2\sum_{j}\intop_{l}ds\intop_{0}^{\infty}dE\intop_{-\pi/2}^{\pi/2}d\theta u_{j}(E)\cos\theta\tau_{i}\left(E,s,\theta\right),
\end{equation}
where $\tau_{i}(E,s,\theta)$ is the conditional probability for an
electron with energy $E$ that enters at position $s$ with an angle
$\theta$ and reaches the boundary of reservoir $C_{i}$ after traversing
the central region. In the context of purely Hamiltonian dynamics,
this probability equals either 1 or 0 \textcolor{black}{\citep{stark2014classical}}.
For the purpose of reaching a concise expression for the net current
$J_{i}\equiv J_{i}^{+}-J_{i}^{-}$ leaving reservoir $C_{i}$, the
transmission coefficient

\begin{equation}
\mathscr{\mathit{T}}_{ji}(E)\equiv\int_{l}ds\int_{-\pi/2}^{\pi/2}d\theta\tau_{j}(E_{r},s,\theta)\cos\theta
\end{equation}
is introduced.

From the volume-preserving property of Hamiltonian dynamics and the
Poincaré-Cartan theorem, it can be demonstrated that \textcolor{black}{\citep{stark2014classical}}

\begin{equation}
\sum_{i}\mathscr{\mathit{T}}_{ji}(E)=\sum_{j}\mathscr{\mathit{T}}_{ji}(E)=2l.
\end{equation}

By combining (3), (4), and (6), the net current out of reservoir $C_{i}$

\begin{equation}
J_{i}=2\sum_{j}\int_{0}^{\infty}dE\mathscr{\mathit{T}}_{ij}(E)\left[u_{i}(E)-u_{j}(E)\right].
\end{equation}
In a similar manner, the net heat flux leaving reservoir $C_{i}$
is calculated by

\begin{equation}
Q_{i}=2\sum_{j}\int_{0}^{\infty}dE\mathscr{\mathit{T}}_{ij}(E)\left(E-\mu_{i}\right)\left[u_{i}(E)-u_{j}(E)\right].
\end{equation}

The entropy production rate of the engine at steady state is expressed
as 
\begin{equation}
\dot{S}\equiv\sum_{i}Q_{i}/T_{i}\text{.}
\end{equation}
\textcolor{black}{To ensure thermodynamic consistency, $\dot{S}$
must be non-negative.}\textcolor{red}{{} }

For a Nernst engine, reservoir $C_{3}$ possesses a higher temperature
compared to reservoir $C_{1}$ ( $T_{3}>T_{1}$), while reservoir
$C_{2}$ has a higher chemical potential than reservoir $C_{4}$ ($\mu_{2}>\mu_{4}$).
Simultaneously, the constraint equations 
\begin{equation}
J_{1}=J_{3}=0\quad\text{ and }\quad Q_{2}=Q_{4}=0,
\end{equation}
are required. These conditions ensure that electron current only occurs
horizontally and heat flow only takes place vertically, as depicted
in Figure 1.

In the following steps, we will explicitly calculate the transmission
coefficients $\mathscr{\mathit{T}}_{ij}(E)$ under the influence of
a strong magnetic field. An electron of energy $E$ moves in a circular
trajectory inside the central region with a radius
\begin{equation}
r(E)=\sqrt{2mE}/(eB)
\end{equation}
because of the Lorentz force. After traveling a distance $\Delta s$
along the boundary (as shown in Fig. 2), the electron eventually collides
with the boundary. In the strong field limit, the radius $r(E)$ of
the electron trajectory is significantly smaller compared to the radius
of the central region. Mathematically, we have $r(E)\ll R$ for the
majority of electrons. As a result, the boundary can be approximated
as a straight line, as illustrated in Fig. 2 . The geometric analysis
demonstrates that
\begin{equation}
\Delta s=2r(E)\cos\theta.
\end{equation}
Since $\Delta s\ll R$, electrons emitted from reservoir $C_{i}$
will either pass to the adjacent reservoir $C_{i+1}$ or return to
$C_{i}$. In other words, electron transmission only occurs between
neighboring reservoirs. Therefore, the transmission coefficient $T_{ji}(E)=0$
for $j\neq i,i+1$. By applying the sum rules given in Eq. (6), it
is recognizes that $T_{ii}(E)=2l-T_{(i+1)i}(E)$. Therefore, we are
now tasked with calculating the transmission coefficient $T_{(i+1)i}(E)$
for the transition from reservoir $C_{i}$ to $C_{i+1}$. To determine
$T_{(i+1)i}(E)$, one should refer to Fig. 2 and observes that a electron
injected from reservoir $C_{i}$ at a specific position $s$ can reach
reservoir $C_{i+1}$ only if $\Delta s\ge s_{i}-s$, where\textcolor{red}{{}
}\textcolor{black}{$s_{i}$} denotes the contact point between reservoir
$C_{i}$ and $C_{i+1}$. By utilizing Eq. (12), this transmission
condition is then given by $\theta_{-}<\theta<\theta_{+}$, where
$\theta_{\pm}=\pm\arccos\left[\left(s_{i}-s\right)/(2r(E))\right]$.
Finally, Eq. (5) can be rewritten as
\begin{equation}
T_{(i+1)i}(E)=\int_{s_{i}-2r(E)}^{s_{i}}ds\int_{\theta_{-}}^{\theta_{+}}d\theta\cos\theta=\pi r(E).
\end{equation}
In the meanwile, the coefficient
\begin{equation}
T_{ii}(E)=2l-\pi r(E)=\pi[R-r(E)].
\end{equation}

\begin{widetext}
By combing Eqs. \textcolor{black}{(4), (6), (13), and (14)}, the analytial
solution of the steady-state current $J_{i}^{-}$ flowing into $C_{i}$
is calculated as follows
\begin{align}
J_{i}^{-} & =2\int_{0}^{\infty}u_{i}(E)\pi\left[R-r(E)\right]dE+2\int_{0}^{\infty}u_{i-1}(E)\pi r(E)dE\nonumber \\
 & =\frac{\sqrt{2\pi m}\pi Re^{\beta_{i}\mu_{i}}}{h^{2}\beta_{i}^{3/2}}-\frac{4\pi me^{\beta_{i}\mu_{i}}}{eBh^{2}\beta_{i}^{2}}+\frac{4\pi me^{\beta_{i-1}\mu_{i-1}}}{eBh^{2}\beta_{i-1}^{2}}.
\end{align}
Equtions \textcolor{black}{(3), (7), and (15)} yield the net current
out of reservoir $C_{i}$
\begin{equation}
J_{i}=\frac{4\pi me^{\beta_{i}\mu_{i}}}{eBh^{2}\beta_{i}^{2}}-\frac{4\pi me^{\beta_{i-1}\mu_{i-1}}}{eBh^{2}\beta_{i-1}^{2}}.
\end{equation}
Through an analogous calculation, the net heat flux leaving reservoir
$C_{i}$ in Eq. (8) is simplified as

\begin{equation}
Q_{i}=\frac{8\pi me^{\beta_{i}\mu_{i}}}{eBh^{2}\beta_{i}^{3}}-\frac{8\pi me^{\beta_{i-1}\mu_{i-1}}}{eBh^{2}\beta_{i-1}^{3}}-\mu_{i}\frac{4\pi me^{\beta_{i}\mu_{i}}}{eBh^{2}\beta_{i}^{2}}+\mu_{i-1}\frac{4\pi me^{\beta_{i-1}\mu_{i-1}}}{eBh^{2}\beta_{i-1}^{2}}.
\end{equation}
\end{widetext}

In the aforementioned circumstances, a noticeable movement of electrons
only occurs between $C_{2}$ and $C_{4}$, whereas the net flow of
heat is solely observed from $C_{3}$ to $C_{1}$. The spontaneous
directed flow of heat leads to the migration of electrons from the
reservoir with lower chemical potential to the reservoir with higher
chemical potential. The power output of the engine is defined as\textcolor{black}{
\begin{equation}
\ensuremath{\ensuremath{P=\left(\mu_{2}-\mu_{4}\right)J_{4}}},
\end{equation}
}while the energy conversion efficiency is given by 

\textcolor{black}{
\begin{equation}
\ensuremath{\ensuremath{\eta=\left(\mu_{2}-\mu_{4}\right)J_{4}/Q_{3}}}.
\end{equation}
}

\begin{figure}
\includegraphics[scale=0.15]{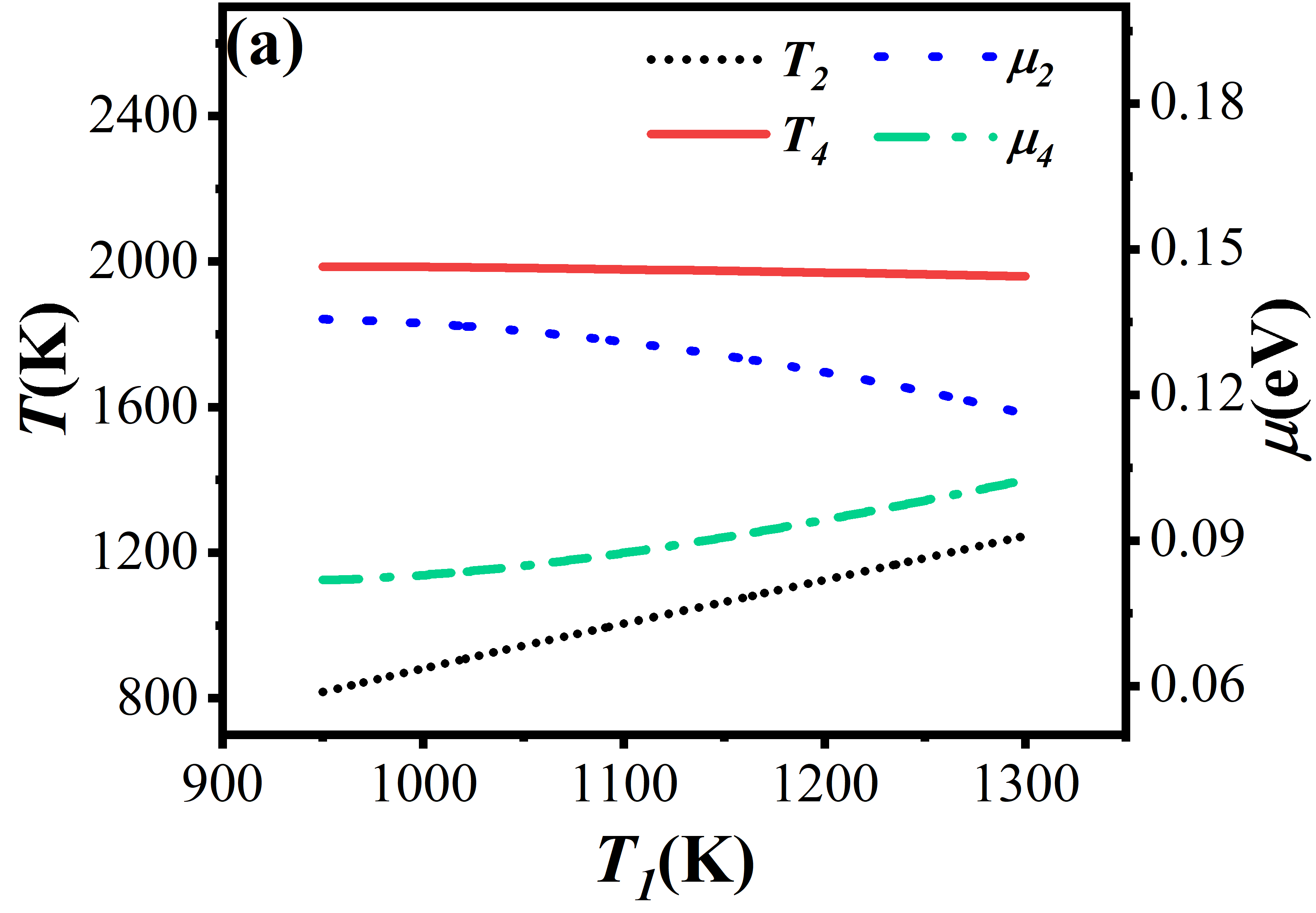} \includegraphics[scale=0.15]{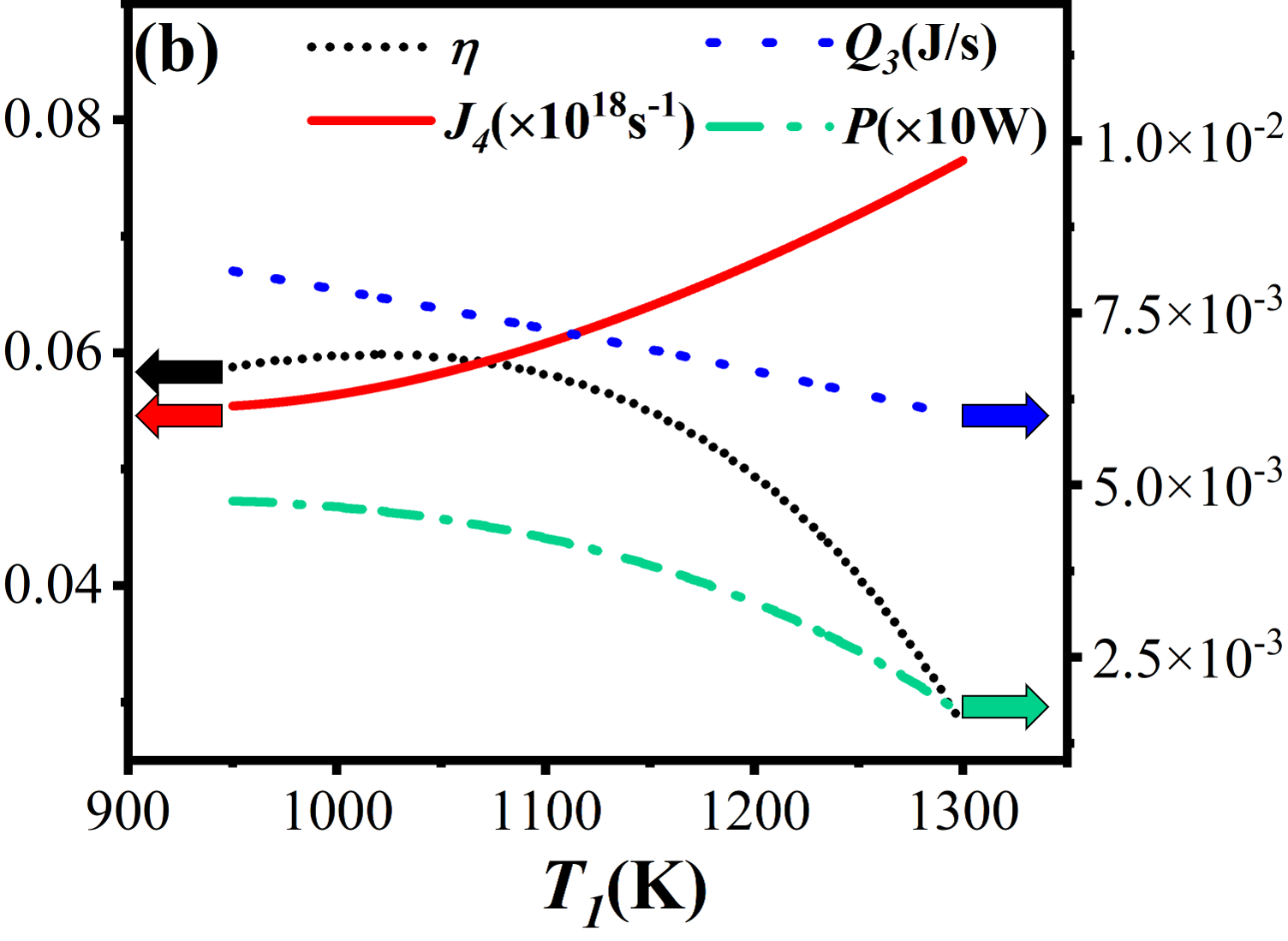}

\includegraphics[scale=0.16]{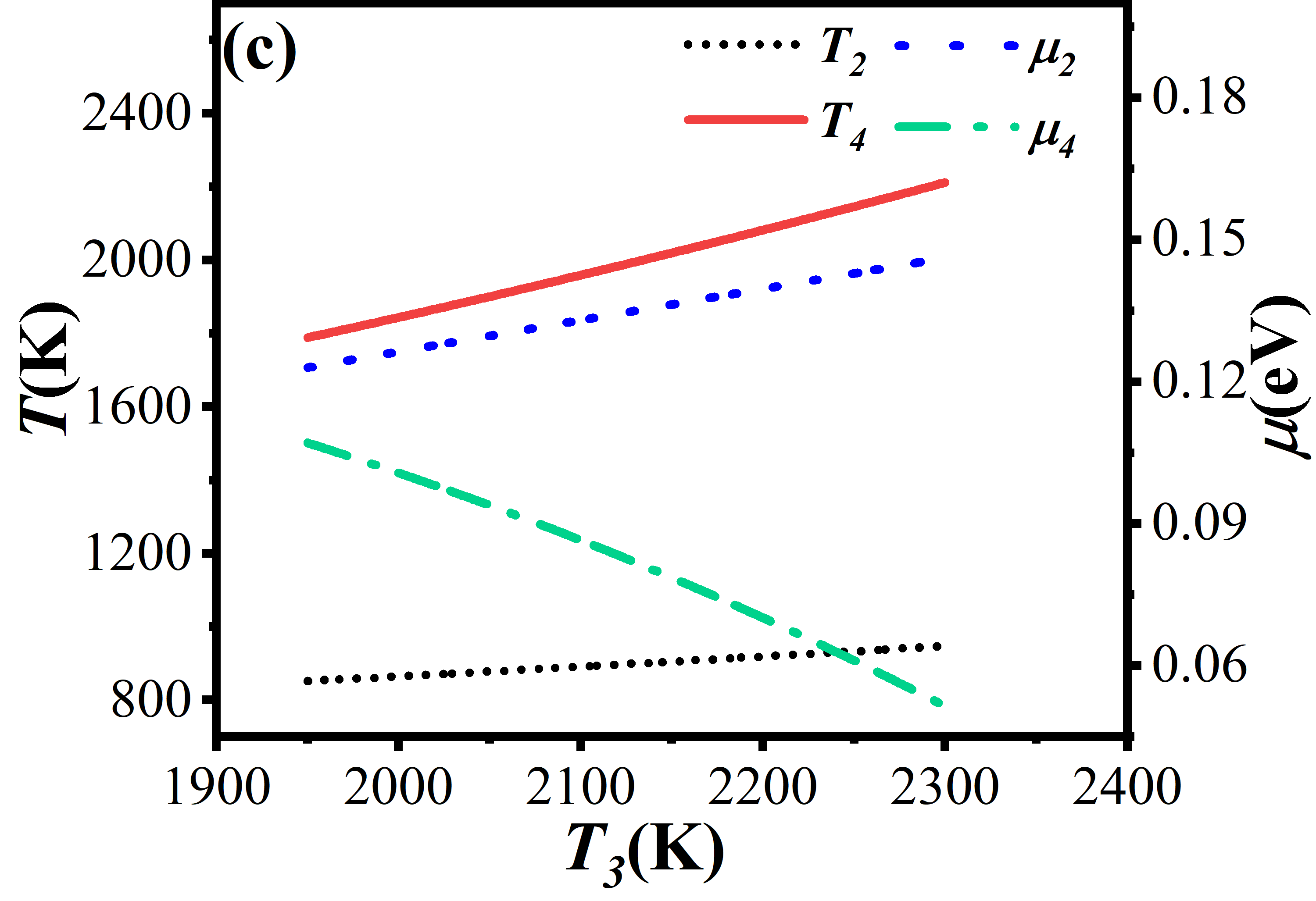} \includegraphics[scale=0.15]{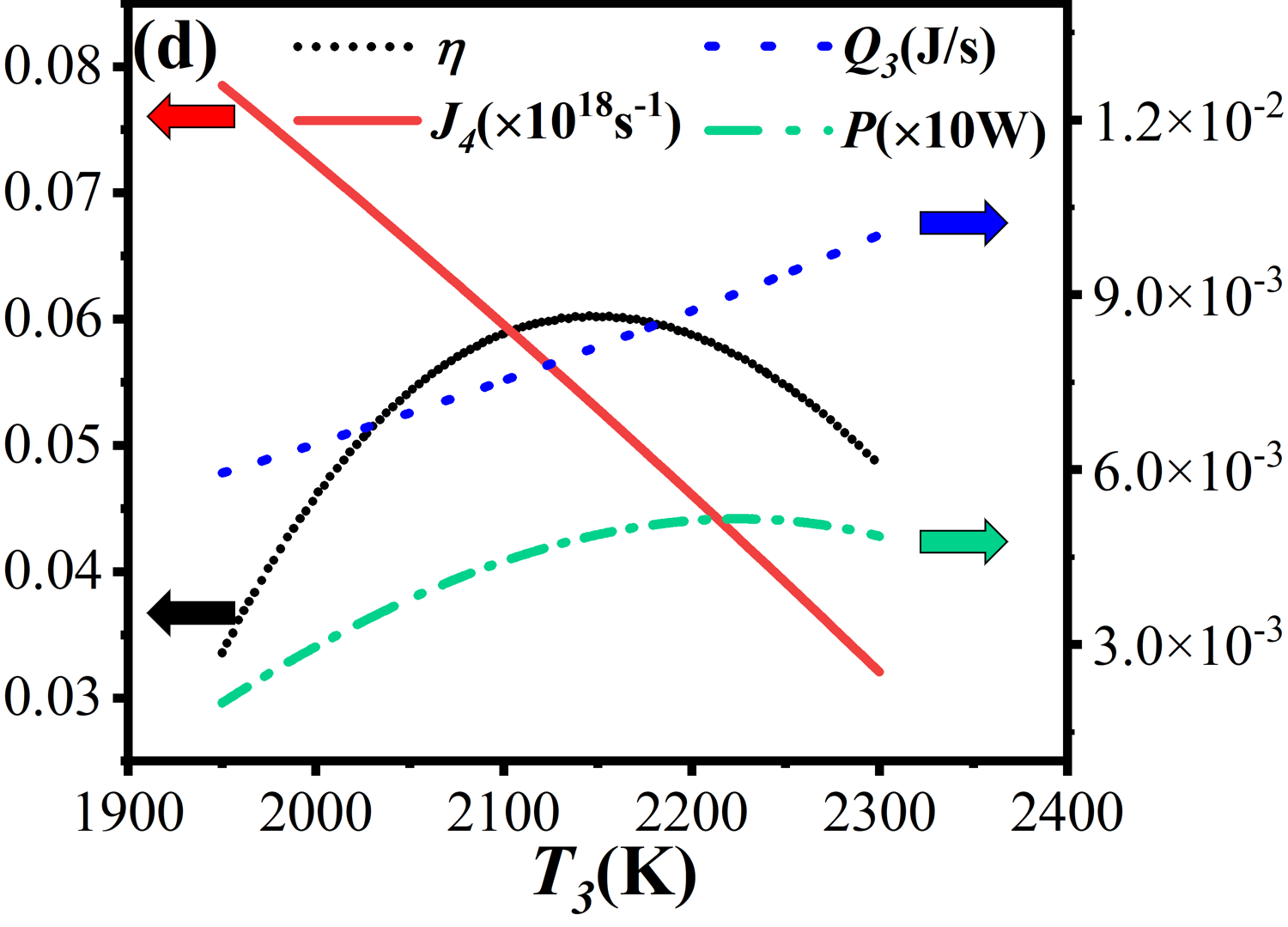}

\caption{(a) The dependence of parameters $\ensuremath{T_{2}}$, $\ensuremath{T_{4}}$,
$\ensuremath{\mu_{2}}$ and $\ensuremath{\mu_{4}}$ of reservoir $C_{2}$
and $C_{4}$, and (b) the efficiency $\eta$, power $P$, current
$J_{4}$, and heat flux $Q_{3}$ on the temperature $\ensuremath{T_{1}}$
of reservoir $C_{1}$, where the temperature $T_{3}$= 2121$K$. (c)
The dependence of parameters $\ensuremath{T_{2}}$, $\ensuremath{T_{4}}$,
$\ensuremath{\mu_{2}}$ and $\ensuremath{\mu_{4}}$ of reservoir $C_{2}$
and $C_{4}$, and (d) the efficiency $\eta$, power $P$, current
$J_{4}$, and heat flux $Q_{3}$ on the temperature $\ensuremath{T_{3}}$
of reservoir $C_{3}$, where the temperature $T_{1}$=1012$K$. The
other parameters $\ensuremath{\mu_{1}}=0.16eV,\ensuremath{\mu_{3}}=0.003eV$,
$R=1m$, and $B=1T$. The arrows in Figs. (b) and (d) indicate the
values of the corresponding physical quantities, which are shown in
the same color.}
\end{figure}

The parameters $\ensuremath{T_{2}}$, $\ensuremath{T_{4}}$, $\ensuremath{\mu_{2}}$
and $\ensuremath{\mu_{4}}$ of reservoir $C_{2}$ and $C_{4}$ can
be determined by utilizing the constraint equations given in Eq. (10).
Figure 3(a) reveals that the temperature $\ensuremath{T_{4}}$ of
reservoir $C_{4}$ remains approximately constant at around 2000K.
As $\ensuremath{T_{1}}$ increases, $\ensuremath{T_{4}}$ shows a
slight downward trend, while the temperature $\ensuremath{T_{2}}$
of reservoir $C_{2}$ exhibits a slightly steeper upward trend. The
behaviors of $\ensuremath{\mu_{2}}$ and $\ensuremath{\mu_{4}}$ are
quite contrasting, where $\ensuremath{\mu_{2}}$ decreases and $\ensuremath{\mu_{4}}$
increases with the increase of $\ensuremath{T_{1}}$. Figure 3(b)
shows that the efficiency $\eta$ reaches a peak value of 5.98\% at
$T_{1}=1023.6K$. As $\ensuremath{T_{1}}$ increases, both the power
$P$ and the heat flux $Q_{3}$ heat experiences a decline. 

According to the expressions in Eqs. (16) and (17), the constraint
equations $J_{1}=0$ and $Q_{4}=0$ can be derived

\begin{equation}
T_{1}^{2}e^{\beta_{1}\mu_{1}}=T_{4}^{2}e^{\beta_{4}\mu_{4}},
\end{equation}

\begin{equation}
\left(2T_{4}^{3}k_{B}-\mu_{4}T_{4}^{2}\right)e^{\beta_{4}\mu_{4}}=\left(2T_{3}^{3}k_{B}-\mu_{3}T_{3}^{2}\right)e^{\beta_{3}\mu_{3}}.
\end{equation}
When $\mu_{1}$ is a given value, the left-hand side of Eq. (20) is
only a function of $T_{1}$. By taking its derivative, it can be found
that within the selected temperature range of $T_{1}$, the derivative
of the left-hand side of Eq. (20) is greater than 0. Due to the constraint
$J_{1}=0$, the value of the right-hand side of Eq. (20) also needs
to be increased accordingly as $T_{1}$ increases. Therefore, $\ensuremath{\mu_{4}}$
increases with the increase of $\ensuremath{T_{1}}$ in Fig. 3(a).
Since the values of $T_{3}$ and $\mu_{3}$ are both given, the right-hand
side of Eq. (21) is a fixed value. Dividing Eq. (21) by Eq. (20),
we can get

\begin{equation}
2T_{4}k_{B}-\mu_{4}=\frac{\left(2T_{3}^{3}k_{B}-\mu_{3}T_{3}^{2}\right)e^{\beta_{3}\mu_{3}}}{T_{1}^{2}e^{\beta_{1}\mu_{1}}}.
\end{equation}
It can be seen that left-hand side of Eq. (22) should decrease with
the increase of $T_{1}$ for the condition $Q_{4}=0$. In Fig. 3(a),
the decrease of $T_{4}$ and the increase of $\mu_{4}$ reflect this
process. Using the same analysis, we can obtain the changes in $T_{2}$
and $\mu_{2}$. As shown in Fig. 3(a), the difference between $\mu_{2}$
and $\mu_{4}$ is a decreasing function of $T_{1}$. For the Nernst
heat engine, this trend reduces the efficiency. However, the increase
of $J_{4}$ and the decrease of $Q_{3}$ shown in Fig. 3(b) serve
to increase the efficiency of the heat engine. For these two reasons,
the efficiency of the heat engine will reach an extreme value. 

The above analysis can also be applied to reveal the dependence of
parameters $\ensuremath{T_{2}}$, $\ensuremath{T_{4}}$, $\ensuremath{\mu_{2}}$
and $\ensuremath{\mu_{4}}$ of reservoir $C_{2}$ and $C_{4}$ on
the temperature $\ensuremath{T_{3}}$ in Fig. 3 (c), as well as the
dependence of the efficiency $\eta$, power $P$, current $J_{4}$,
and heat flux $Q_{3}$ on the temperature $\ensuremath{T_{3}}$ in
Fig. 3(d). 

\begin{figure}
\includegraphics[scale=0.2]{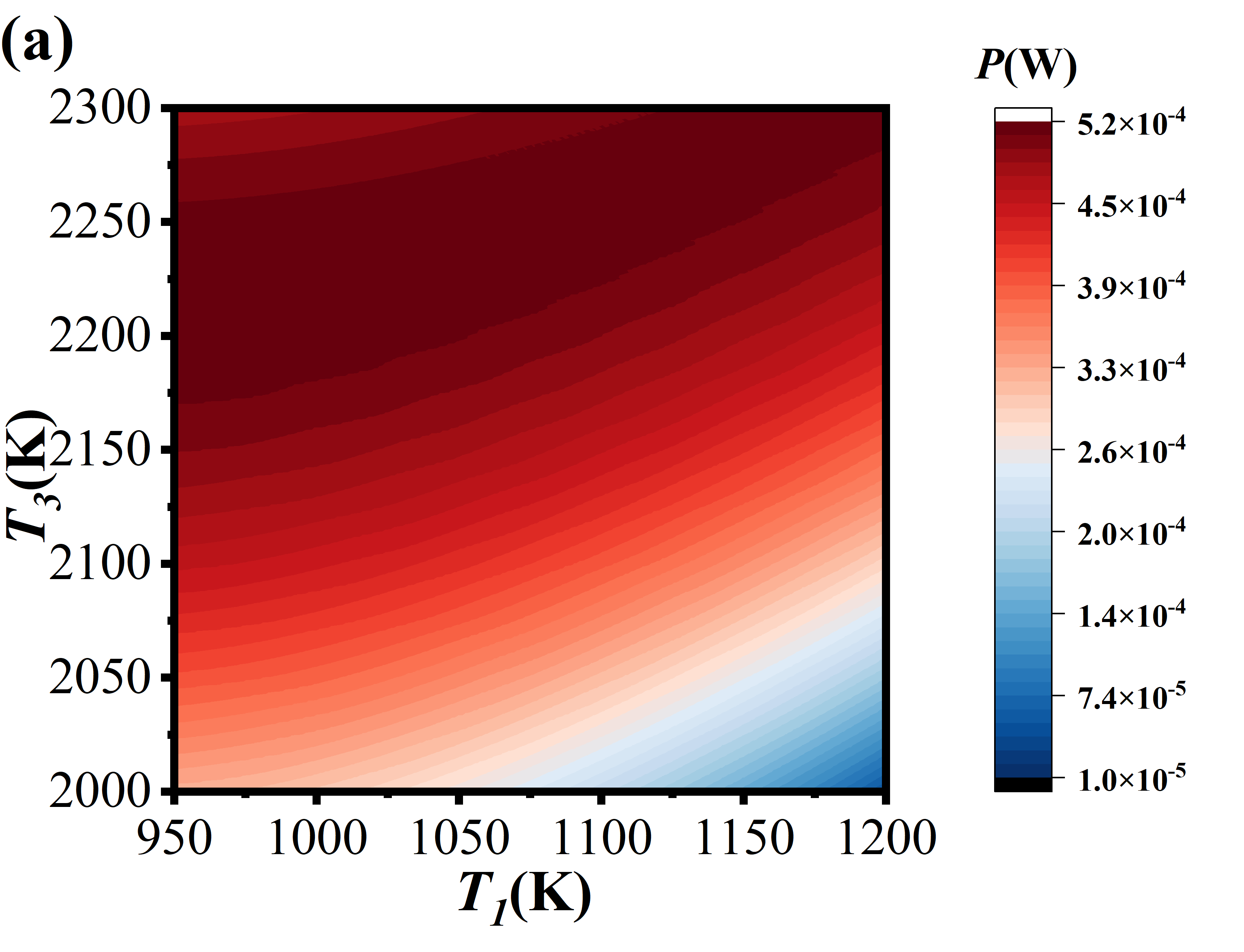}

\includegraphics[scale=0.2]{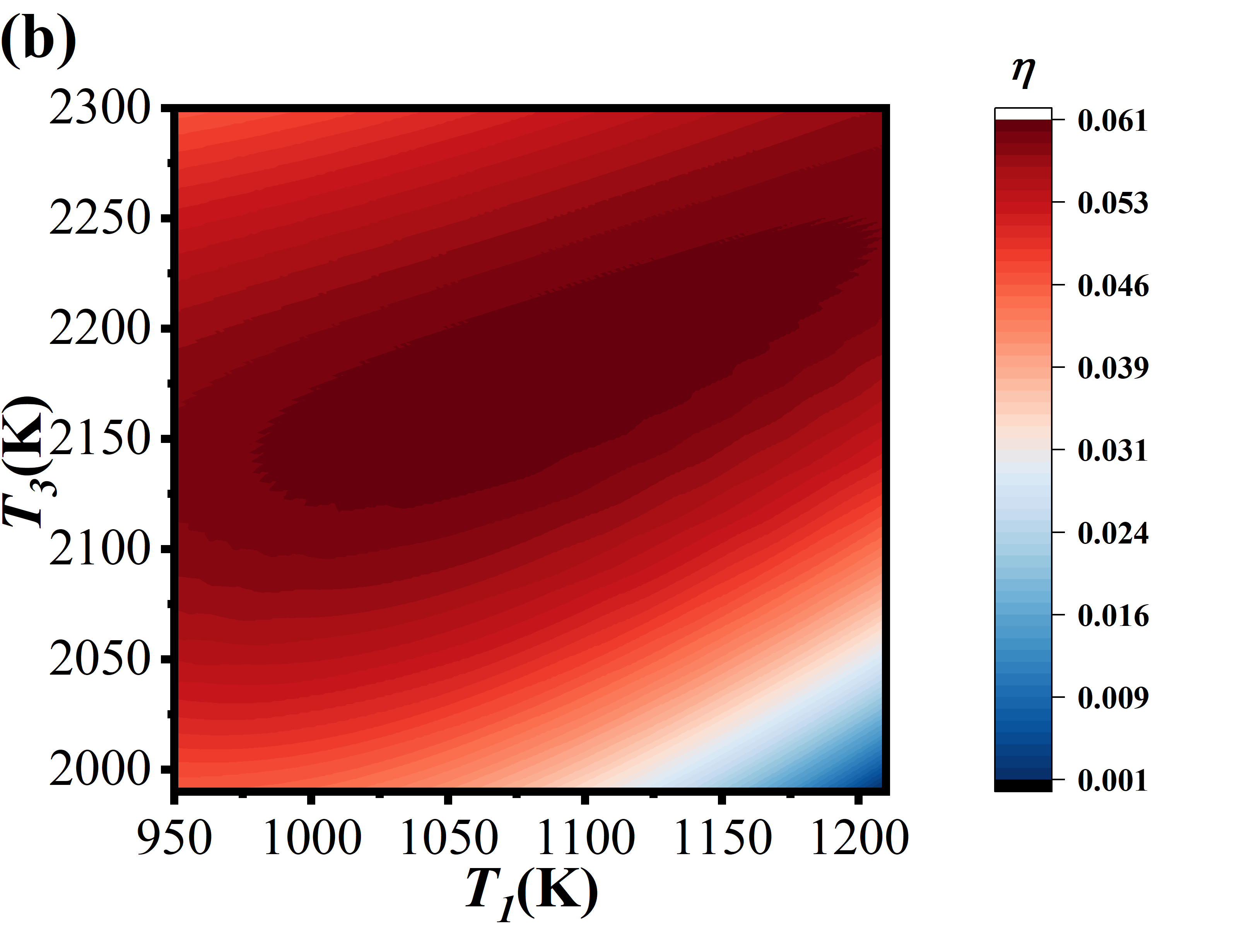}

\caption{The two-dimensional graph of (a) the power $P$ and (b) efficiency
$\eta$ varying with $\ensuremath{T_{1}}$ and $\ensuremath{T_{3}}$,
while keeping the other parameters the same as those used in Fig.
3.}
\end{figure}

The relationship between the power $P$ and efficiency $\eta$ in
relation to temperatures $T_{1}$ and $T_{3}$ is illustrated in Figure
4(a). When the given value of $T_{3}$ is small, $P$ decreases significantly
as $T_{1}$ increases. However, when the given value of $T_{3}$ is
large, $P$ starts to increase as $T_{1}$ increases. When $T_{1}$
is held at a constant value, the power $P$ exhibits an extremum as
$T_{3}$ varies. Figure 4(b) depicts a region where the efficiency
reaches its maximum value as both $T_{1}$ and $T_{3}$ vary. Optimal
performance, characterized by enhanced power and efficiency, is attained
when both temperatures fall within the dark red area of the contour
plot.

In this work, we perform numerical simulations on a Nernst-based thermionic
engine. By specifying the temperature and chemical potential of certain
heat sources, we calculate the resulting changes in temperature and
chemical potential of unknown heat reservoirs. Moreover, we determine
the system's power output and the heat flux that drives the system.
Surprisingly, we discover that by determining the temperature or chemical
potential of $C_{1}$\textcolor{red}{{} }\textcolor{black}{and}\textcolor{red}{{}
}$C_{3}$ and optimizing the remaining parameters, we can achieve
the maximum power and efficiency of the Nernst heat engine. This discovery
highlights the practicality of optimizing the performance of the Nernst
heat engine.

This work has been supported by the National Natural Science Foundation
(12075197 and 12364008), Natural Science Foundation of Fujian Province
( 2023J01006), Educational Teaching Reform Research Project of University
Physics Discipline Alliance of Fujian Province (FJPHYS-2023-A02),
and Fundamental Research Fund for the Central Universities (20720240145). 

\bibliographystyle{apsrev4-1}
\bibliography{Nernst}

\end{document}